# DSTN (Distributed Sleep Transistor Network) for Low Power Programmable Logic array Design

Pradeep Singla
Deptt. Of Elect. & Comm. Engg.
Hindu College of Engineering
Sonipat- India

Kamya Dhingra
Deptt. Of Elect. & Comm. Engg.
B.M. Instt. of Engg. & Mgmt.
Sonipat- India

Naveen Kr. Malik
Deptt. Of Elect. & Comm. Engg.
Hindu College of Engineering
Sonipat- India

## ABSTRACT
With the high demand of the portable electronic products, Low- power design of VLSI circuits & Power dissipation has been recognized as a challenging technology in the recent years. PLA (Programming logic array) is one of the important off shelf part in the industrial application. This paper describes the new design of PLA using power gating structure sleep transistor at circuit level implementation for the low power applications. The important part of the power gating design i.e. header and footer switch selection is also describes in the paper. The simulating results of the proposed architecture of the new PLA is shown and compared with the conventional PLA. This paper clearly shows the optimization in the reduction of power dissipation in the new design implementation of the PLA. The transient response of the power gates structure of PLA is also illustrate in the paper by using TINA-PRO software.

## Keywords
Low Power design, Sleep Transistor, Header, Footer, Power gating, TINA- PRO, PLA

## 1. INTRODUCTION
The advance laptop computers, digital cameras, processors, wearable computers, smart cards etc. are the witness of rapid growth of the semiconductor industries since the last decades [5]. This rapid and explosive growth forces designer to struggle for the smaller silicon area, longer battery life and more reliability. So, the demands for the low power VLSI systems are also increasing for designing such a low power devices [9]. But the power dissipation is one of the major concerning factors in case of low power VLSI design. In case of CMOS (Complementary metal oxide semiconductor) design of the circuits, both PMOS and NMOS are contribute equally to the circuit operation & the basic source of power dissipation in the digital circuits are following [11].

$$P_{avg.} = P_s + P_{s.c.} + P_{leakage} = \alpha_{0\to1} C_l . V^2_{dd} . f_{clk} + I_{s.c.} . V_{dd} + I_{leakage} . V_{dd}$$

Where $C_l$ is the load capacitance, $f_{clk}$ is the clock frequency, $\alpha_{0\to1}$ is the probability that a power consuming when transistion occurs, $I_{s.c.}$ is the short circuit current (when both nmos and pmos active), $I_{leakage}$ is the leakage current arises from subthreshold effects.So, as we improving the technology or scaling the technology, the leakage power is also increasing exponentially during the standby mode or at time of inactivity.[4] There are numbers of methods proposed in the literature for reducing the sub threshold leakage [1] but power gating is one of the effective methods for reduction of sub threshold leakage [3]. In such a design, multi-threshold CMOS has been introduced with the low & high threshold Voltage ( $V_{th}$ ) module connected to the ground and to power supply respectively caled sleep transistors as shown in fig.1

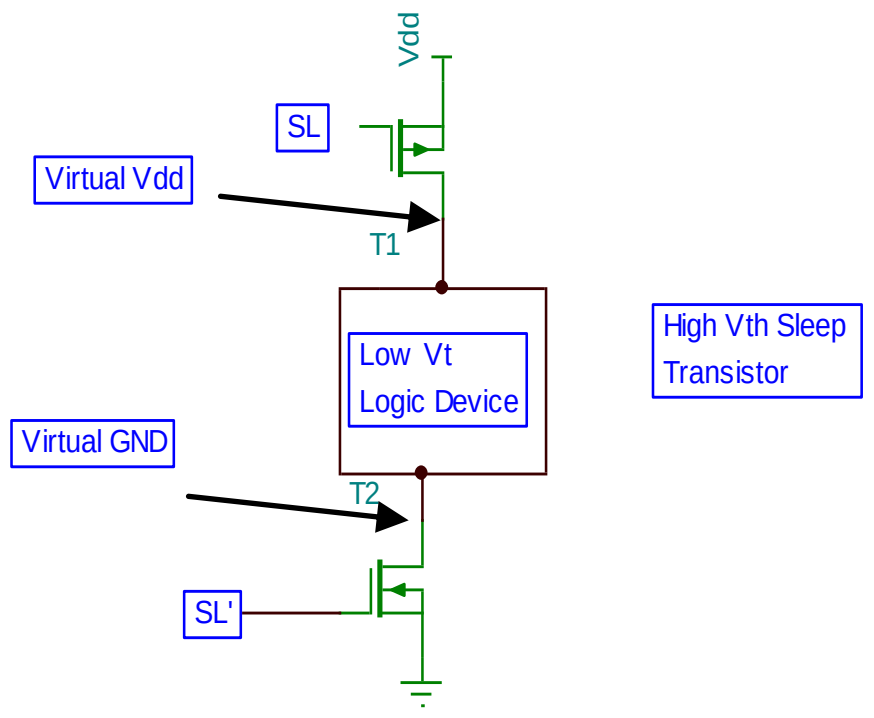

**Fig.1 A Power Gating structure**

The active and sleep are the two operational modes of the multi- threshold CMOS technology for the power saving. In this technique, the low threshold voltage logic device from power supply and the ground via sleep transistor is also known as power gating. In this paper, by seeing the benefits of the PLA (Programmable logic array) structure over the Non-PLA structure, we propose a power efficient design of programmable logic array for the low power industrial applications by using the multi threshold distributed sleep transistor network. The programmable logic array has many applications in the medical and industrial field. In order to demonstrate the proposed architecture of power efficient PLA, a 3- input PLA which can perform any $2^3$ functions using the combination of 8 min terms is also designed and compare the result with the conventional PLA by using the simulator TINA-PRO. This structure has a benefit over the non-PLA structure because of its programmable facility [9]. This proposed low power PLA can be used to implemented numbers of Boolean functions like adder/subtractor and it can be changed to perform desired function.

## 2. OVERVIEW OF SLEEP TRANSISTOR
A sleep transistor may be of PMOS or NMOS with high threshold voltage ($V_{th}$) transistor which is placed in series with a low $V_{th}$ device module. In the power gating structure and circuit operates in two different modes

*Active Mode [3]*
In the active mode, the sleep transistor is turned ON and acts as the functional redundant resistance.
*Sleep Mode [3]*
In this mode, the sleep transistor is turned OFF to reduce dynamic and leakage power in the standby mode.
When a sleep transistor is placed near Vdd, then called header switch and when a sleep transistor put near to the ground, called footer switch as shown in fig.2 (a), fig.2 (b).

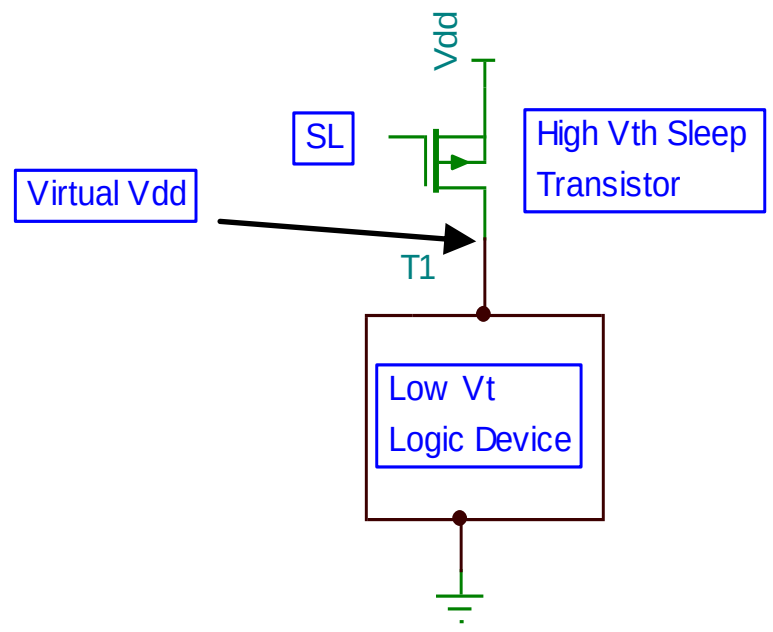

**Fig 2(a) Header Switch**

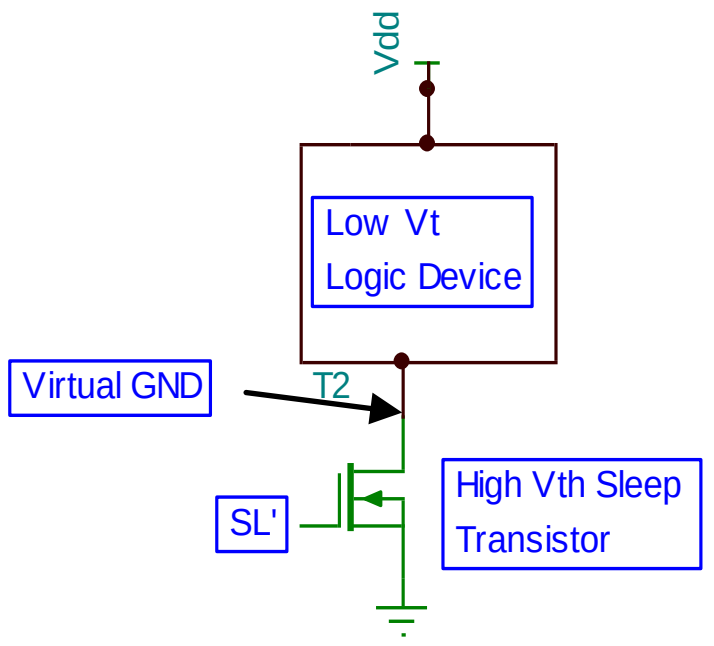

**Fig.2 (b) Footer Switch**

As we can see in the figures, the PMOS is used for implementing the header switch putted near to Vdd to control the supply and the NMOS is putted near to GND to control GND. But PMOS transistor is less leaky than the NMOS transistor and PMOS has lower drive current than the NMOS transistor [2] [7]. Due to this PMOS takes more area than the NMOS which is not an aspect for the very large scale integration. So, by seeing the advantages of the NMOS i.e. footer switch of high drive current & smaller area over the PMOS, we consider the footer switch only in our propose design. In the power gating design, we only used either header or footer switch due to constraint of sub-1V power supply voltage [2].

## 2.1 Analysis of Footer Switch

In this, we are going to setup a relationship between ground voltage and the leakage saving. For this, we assume a footer switch as shown in fig 2(b) under the assumption that this footer is biased in the weak inversion i.e. $V_g < V_{th}$. So, the leakage of the single transistor provides the leakage of the logic circuit [10].
So,

$$I_{leakage}\ (circuit) = I_{leakage}\ (Footer)$$

But

$$I_{leakage} = I_o \left(\frac{W}{L}\right) 10^{\frac{(V_{g} - V_{th}) + \eta(V_{ds})}{SS}} \text{ for the logic circuit} \quad \ldots\ldots\ldots\ldots\ldots(1)$$

So, equation becomes

$$I_o \left(\frac{W\ cicuit}{L}\right) 10^{\frac{(-Vthc)+\eta(V_{dd}-Vgnd)}{SS}} = I_o \left(\frac{WFooter}{L}\right) 10^{\frac{(V_{g} - V_{thF})+\eta(V_{ds})}{SS}} \quad \ldots\ldots\ldots\ldots\ (2)$$

Vthc and VthF shows the threshold voltage of the logic circuit and the footer device resp., $\eta$ is the Drain induced barrier lowering ( a secondary effect) coefficient and ss is the sub threshold slope.
After solving eq 2.

$$V_{gnd} = \frac{-V_g + S_s \log_{10}\left(\frac{Wcircuit}{Wfooter}\right) + (Vthf - Vthc\ + \eta Vdd}{2\eta} \ldots\ldots(3)$$

The equation 3 shows that Vgnd is proportional to the gate voltage Vg with negative slope. So, if the footer gate voltage is increased, there is decrease in the ground potential and vice-versa.
To make the trade-off between leakages saving and wakeup-overhead, the control over Vgnd should be needed. So,

$$\frac{I\ Sleep}{I\ active} = 10^{-\left(\frac{\eta(Vdd - Vgnd)}{Ss}\right)} \ldots..(4)$$

So, by the above terms, higher Vgnd results in higher leakage saving.

# 3. MOTIVATION FOR WORK

This paper introduces the new design of the PLA by using the sleep transistor network. As we know that, the PLA are the standard off shelf part that offers customers a wide variety of logic capacity, sped, & voltage characteristics [9]. The PLA's are used in numbers of application like ultrasonic flow detection, DSP's etc. The PLA are faster than high speed DSP's. Thus we propose a power efficient design of a digital system by combing the effect of PLA's and power gating design.

# 4. PRPOSED IMPLEMENTATION

The PLA consists of AND array and OR arrays shown in fig. 3

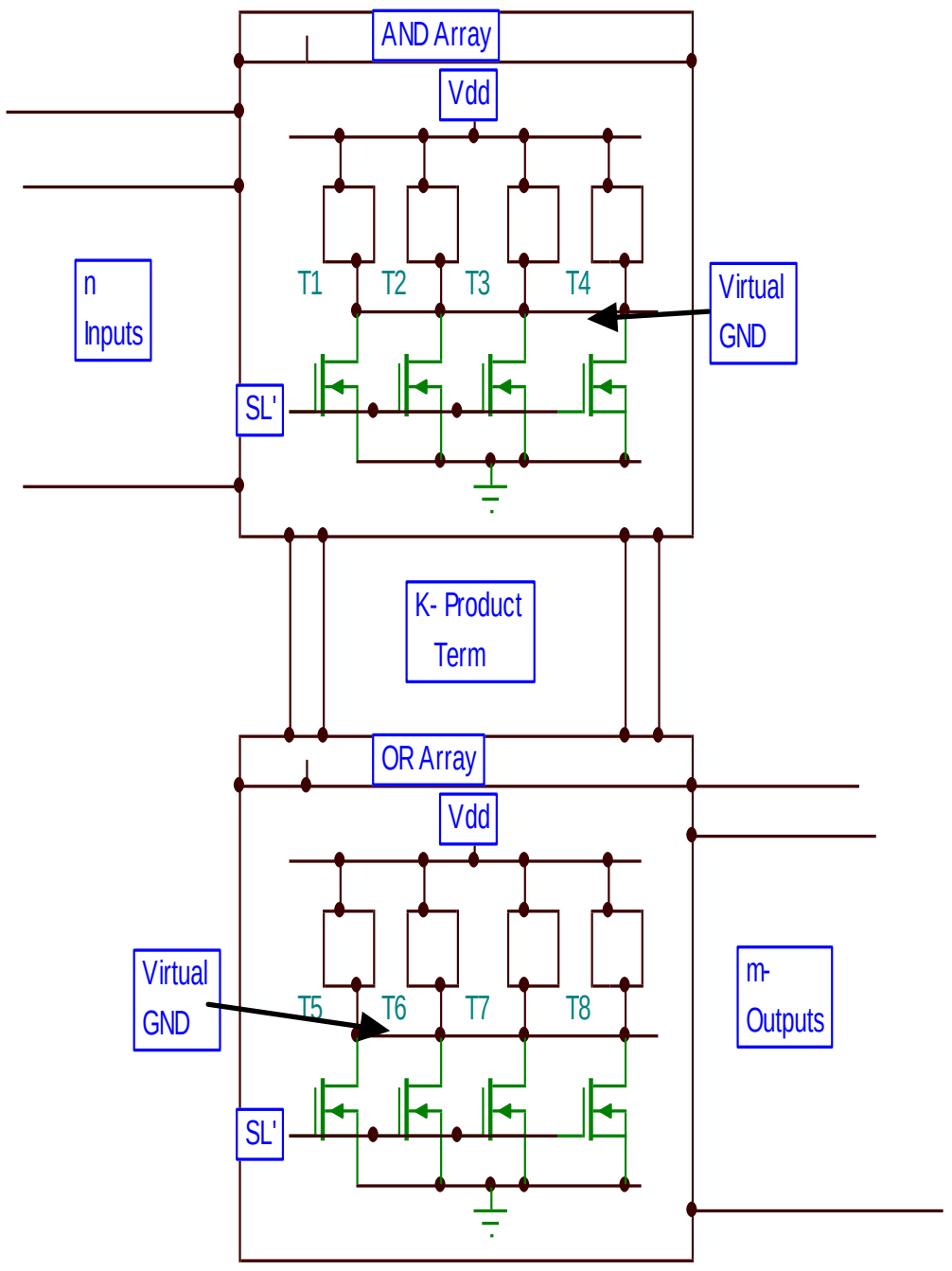

**Fig 3. Proposed PLA with Sleep Transistors**

In this architecture, there are n numbers of inputs provides to the AND array which will provide the product term equals to k. The OR array taking the product term as inputs and provide the sum of product. The internal structure of each array is implemented by sleep transistor with footer configuration. By this architecture we can easily change this structure for getting the desired Boolean function at any time. In this work we are taking the example of 3- input PLA in which the number of inputs provided to the AND array are 3 and this will provide us 8- min terms or product terms. The OR gate will perform for the 8- min terms and provide the sum of those.

## 4.1 Power Gated Design of PLA

For this design we use the TINA-PRO software. TINA Design suit is a powerful yet affordable software package for analyzing, designing and real time testing of analog, digital, VHDL, and mixed electronic circuits and their layouts. For the simplicity we first make the macro for the every AND gate and then connect these macros in a manner as describes in last section.

### *4.1.1 AND Array*

In this array we use the wattmeter for the power dissipation measurement purpose. VF1-VF8 is the output voltages and PM1- PM3 is power measurement wattmeter. The inverter used in the architecture is high speed CMOS inverter. The fig.4 (a) shows the complete structure of AND array whereas the fig.4 (b) illustrate the AND macro used in the PLA structure.

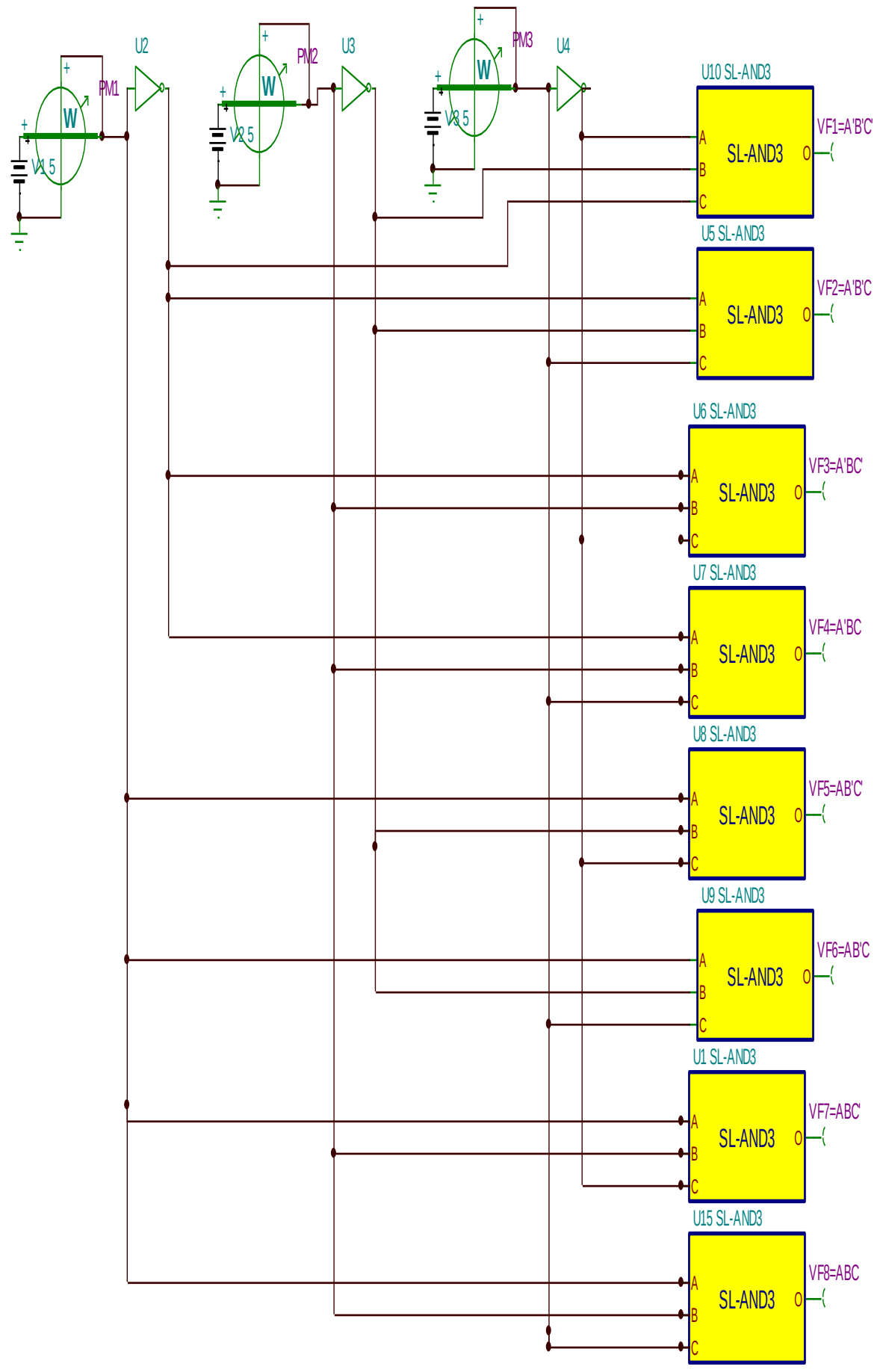

**Fig 4(a) Power measurement set-up architecture of AND Array of PLA with Sleep transistor**

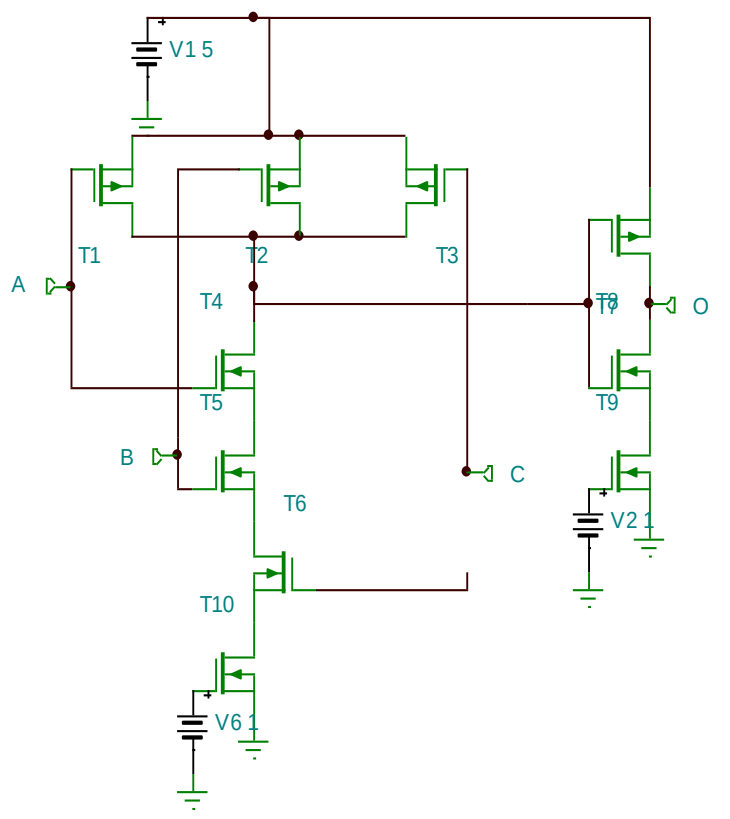

**Fig 4(b) Internal Implementation of SL-AND3 Macro**

### *4.1.2 OR Array*

The fig.5 (a) shows the complete structure of OR array whereas the fig.5 (b) illustrate the OR macro used in the PLA structure.

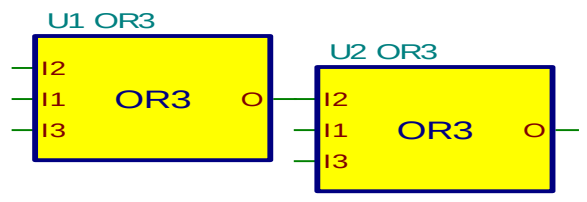

**Fig.5 (a)) Architecture of OR array of PLA with Sleep transistor**

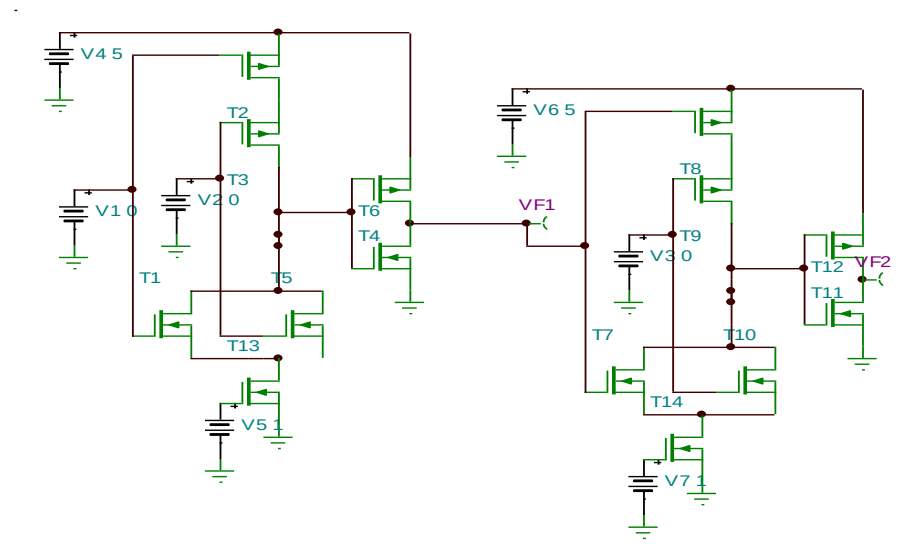

**Fig 5(b) Internal Implementation of OR3 Macro**

## 5. RESULTS

In this section, we describe the performance analysis using TINA- PRO. For analyzing the parameters like power consumption and transient response, we consider the input value at variable A, B, & C are logic 1 or +5V unit step.

### 5.1 Power Analysis

For the power analysis, we take both of the PLA's structure i.e. Conventional PLA and proposed PLA. After supplying the inputs at different levels we get the results as shown in table1.

**Table 1: Power Consumption comparison between Conventional & Proposed PLA**

| Input Vector(v) | Power Consumed (pW) | | | | | |
|---|---|---|---|---|---|---|
| | Conventional PLA | Proposed PLA | Conventional PLA | Proposed PLA | Conventional PLA | Proposed PLA |
| | Line A | Line A | Line B | Line B | Line C | Line C |
| 000 | 0 | 0 | 0 | 0 | 0 | 0 |
| 001 | 0 | 0 | 0 | 0 | 124.98 | 88.34 |
| 010 | 0 | 0 | 100.47 | 88.35 | 0 | 0 |
| 011 | 0 | 0 | 100.89 | 88.75 | 124.98 | 112.35 |
| 100 | 57.87 | 47.31 | 0 | 0 | 0 | 0 |
| 101 | 58.2 | 47.67 | 0 | 0 | 124.98 | 112.38 |
| 110 | 58.1 | 47.61 | 100.26 | 88.17 | 0 | 0 |
| 111 | 58.44 | 47.97 | 100.68 | 88.59 | 124.98 | 112.39 |

### 5.2 Transient Response Analysis

"Transients", a term we'll use for simplicity here, are actually "Transient Voltages". More familiar terms may be "surges" or "spikes". Basically, transients are momentary changes in voltage or current that occurs over a short period of time. The Graph Drawn below is the comparable Transient response of the Conventional PLA and Proposed PLA.

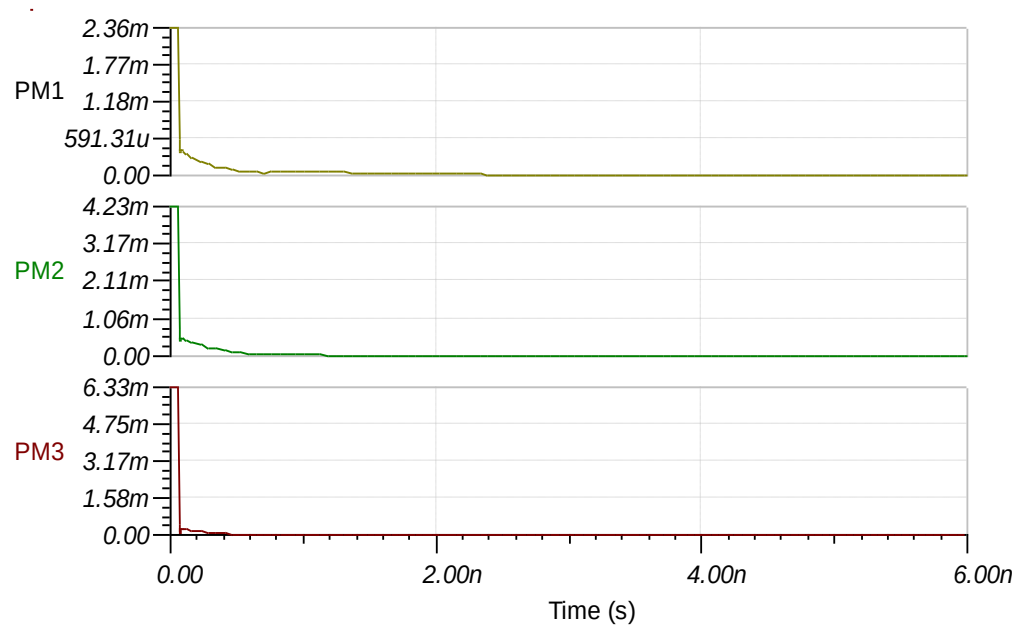

**Fig 6(a) Transient Response of Conventional PLA**

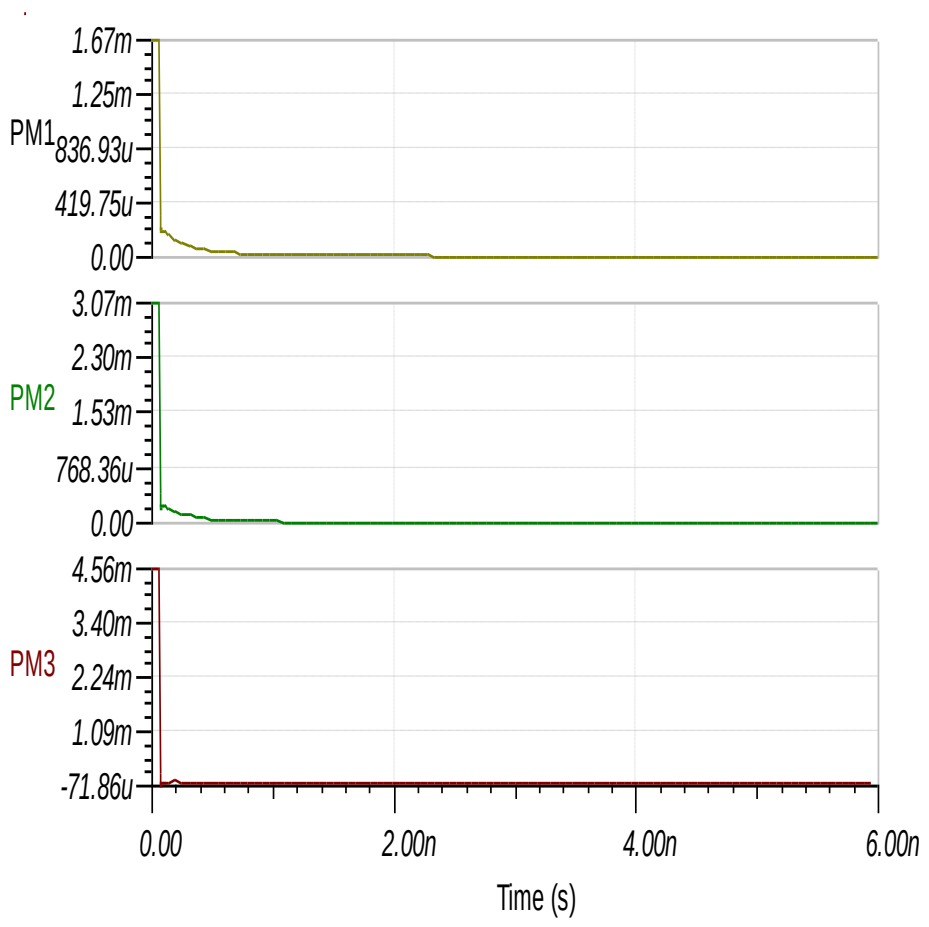

**Fig 6(b) Transient Response of Proposed PLA**

Fig 7 shows the resultantly power consumption comparison between the Conventional & proposed PLA at input logic high on three lines A, B, & C.

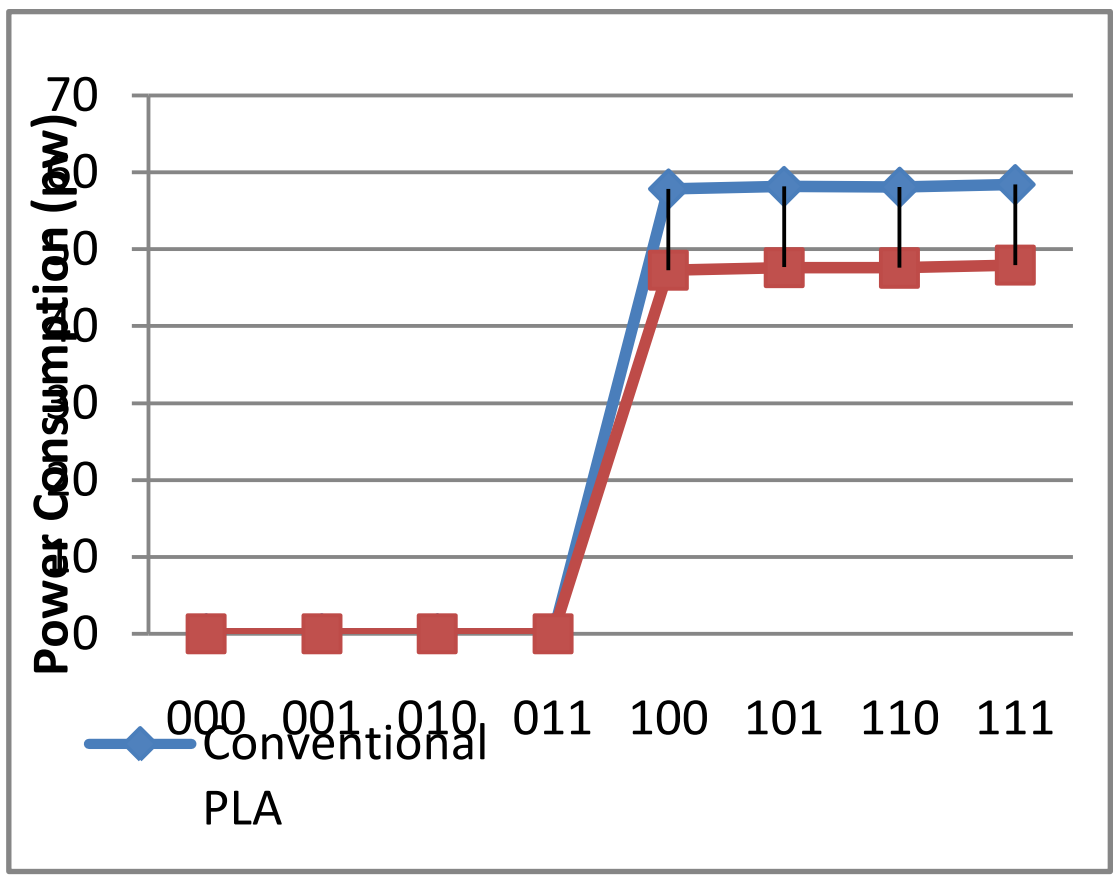

**Fig. 7(a) Average power consumption comparison on Line A**

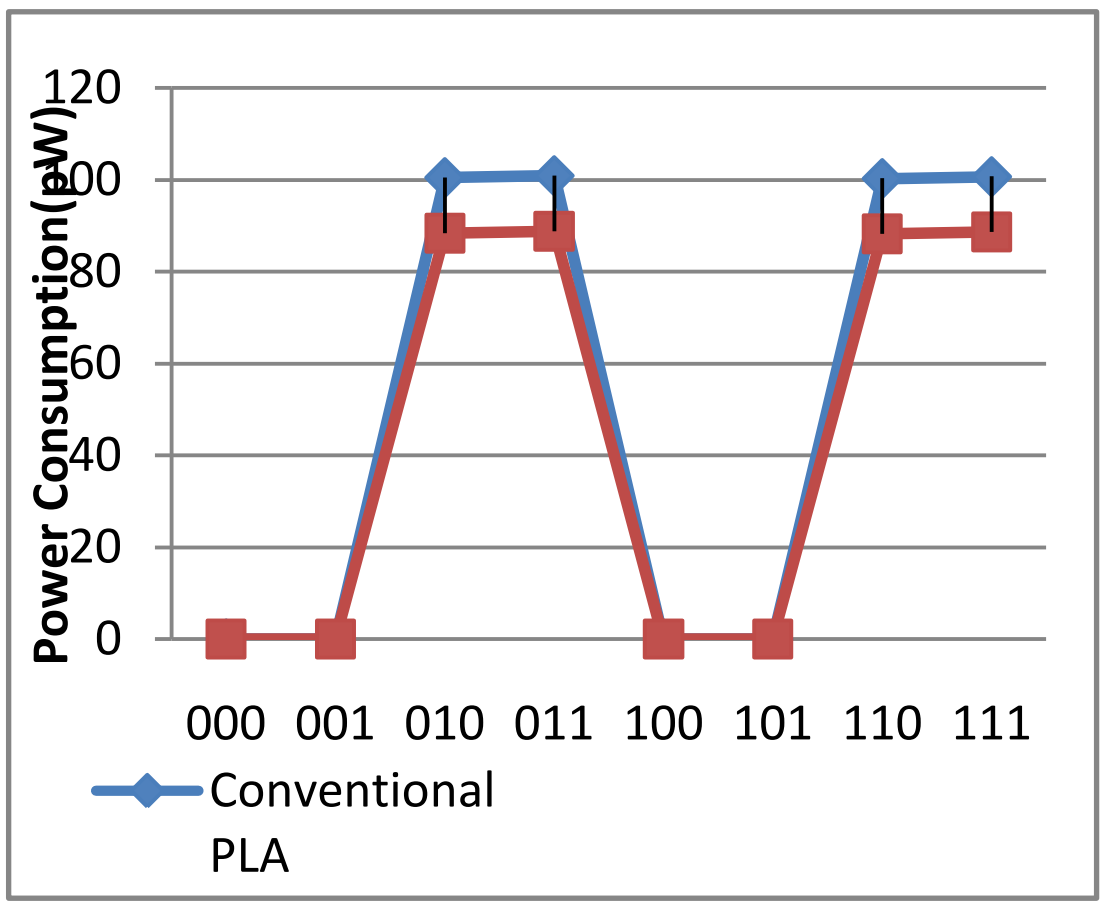

**Fig. 7(b) Average power consumption comparison on Line B**

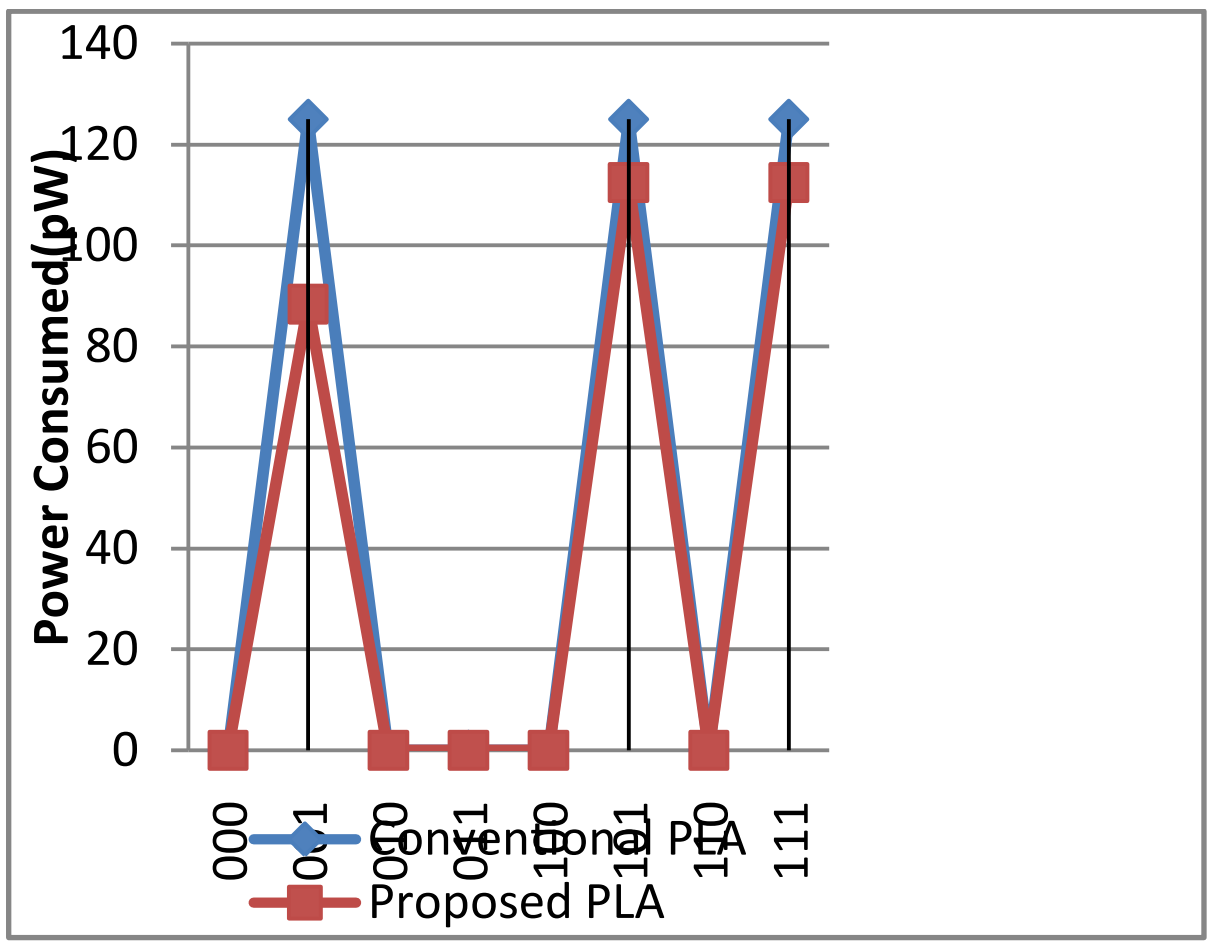

**Fig. 7© Average power consumption comparison on Line C**

## 6. CONCLUSION

In this paper, we emphasis on the efficient physical level designing of the digital hardware for reducing the sub threshold leakage by using the power gating structure. The PLA which is generally used in the industrial applications is designed in this paper by the sleep transistor. The table 1 shows the power consumption comparison results of conventional & proposed PLA and shows the proposed PLA is much efficient than the conventional Plate transient response analysis has been also shown in the paper. This PLA is also used to design number of Boolean function like adder, subtractor etc. and can be changed at any time for desired function. The resultant function will also be power efficient than the conventional ones. So, we can conclude that this approach of design is very efficient for the low power VLSI design and this proposed PLA structure can be used for the low power industrial applications.

## 7. ACKNOWLEDGMENT

Authors wish to Thank God for the wisdom and perseverance that he has been bestowed upon us during this research project, and indeed, throughout our life.

## 8. REFERENCES

[1] Abdullah A, Fallah F, and Pedram M, (Jan 2007) " A robust power gating structure and power mode transition strategy for MTCMOS design" IEEE Trans. Very large Scale integration., vol 15, No 1, pp.80-89.

[2] Howard D., Shi K (2006), "sleep Transistor design and implementation simple concepts yet challenges to be optimum "proc. VLSI-DAT, pp. 1-4.

[3] C.Chrisim Gnana suji, S. Maragatharaj, "Performance analysis of power gating in low power VLSI circuits" Proc. Of ICSCCN, pp. 689-694, 2011IEEE.

[4] Changbo Long and L. He., "Distributed sleep transistor network for power reduction," IEEE Trans. Very large scale integer, vol. 12, no. 9, pp. 937-946, Sep 2004.

[5] Designing Low-Power Circuits: Practical Recipes by Luca Benini Giovanni De Micheli Enrico Macii.

[6] J. P. Uyemura, Introduction to VLSI Circuits and Systems. New York: Wiley, 2002.

[7] C. Long, J. Xiong, and L. He, "On optimal physical synthesis of sleep transistors," ill Proc. ISPD, 2004, pp. 156-161.

[8] Chang H, Lee C, and Sapatnekar S.S, (2005) "Full-chip analysis of leakage power under process variations, including spatial correlations" in Proc. Des. Autom. Coni (DAC), pp. 523-528.

[9] Pradeep Singla and Naveen Kr. Malik, " A cost- effective design of reversible programmable logic array" , International Journal of Computer applications Vol.41(15), pp.41-46, 2012.

[10] S. Mutoh et al., " 1-V power supply high speed digital circuits technology with multihreshold voltage CMOS," JSSC, vol. SC-30,pp. 847-854, aug. 1995.

[11] Umea Normal et al., " A low power high speed adders using MTCMOS Technique" International journal of computational engineering & Management, vol. 13, pp. 65-69, July 2011.